\documentclass[letterpaper, 10 pt, conference]{ieeeconf}  

\IEEEoverridecommandlockouts                              

\usepackage{cite}
\usepackage{amsmath, amssymb, amsfonts, mathrsfs}

\usepackage{graphicx}
\usepackage{textcomp}
\usepackage[dvipsnames]{xcolor}
\usepackage{pgfplots}
\pgfplotsset{compat=newest}
\graphicspath{{figures/}}
\usepackage{multirow}
\usepackage{makecell}

\usepackage{ntheorem}

\usepackage{algorithm}
\usepackage[noEnd=true]{algpseudocodex} 

\usepackage{booktabs}
\usepackage{adjustbox}
\usepackage{hyperref}
\usepackage{optidef}

\usepackage{tikzducks}  

\usepackage{siunitx}    
    
\newcommand{\white}[1]{\textcolor{white}{#1}}

\DeclareMathOperator{\bx}{\boldsymbol{x}}

\DeclareMathOperator{\bu}{\boldsymbol{u}}
\DeclareMathOperator{\bv}{\boldsymbol{v}}
\DeclareMathOperator*{\argmin}{arg\,min}
\DeclareMathOperator{\proj}{proj}
\DeclareMathOperator{\rob}{rob}

\newcommand\copyrighttext{%
	\footnotesize \copyright 2025 IEEE.  Personal use of this material is permitted.  Permission from IEEE must be obtained for all other uses, in any current or future media, including reprinting/republishing this material for advertising or promotional purposes, creating new collective works, for resale or redistribution to servers or lists, or reuse of any copyrighted component of this work in other works.}
\newcommand\copyrightnotice{%
	\begin{tikzpicture}[remember picture,overlay]
		\node[anchor=south,yshift=10pt] at (current page.south) {\fbox{\parbox{\dimexpr\textwidth-\fboxsep-\fboxrule\relax}{\copyrighttext}}};
	\end{tikzpicture}%
}

\title{\LARGE \bf
   Trajectory Planning with Signal Temporal Logic Costs\\ using Deterministic Path Integral Optimization
   }

\author{Patrick Halder$^{1,3,*}$, Hannes Homburger$^{2,*}$, Lothar Kiltz$^{1}$, Johannes Reuter$^{2}$, and Matthias Althoff$^{3}$
\thanks{$^{1}$ZF Friedrichshafen AG, 88046 Friedrichshafen, Germany. \scriptsize \tt \{patrick.halder, lothar.kiltz\}@zf.com}%
\thanks{$^{2}$Institute of System Dynamics, HTWG Konstanz, 78562 Konstanz, Germany. \scriptsize \tt \{hhomburg, jreuter\}@htwg-konstanz.de}%
\thanks{$^{3}$School of Computation, Information and Technology, Technical University of Munich, 85748 Garching, Germany. \scriptsize \tt \{patrick.halder, althoff\}@tum.de}%
\thanks{$^{*}$Both authors contributed equally to the paper.}%
}

\begin{document}

\newpage
	\maketitle
	\thispagestyle{empty}
	\pagestyle{empty}

	\begin{abstract}
    Formulating the intended behavior of a dynamic system can be challenging. Signal temporal logic (STL) is frequently used for this purpose due to its suitability in formalizing comprehensible, modular, and versatile spatio-temporal specifications. Due to scaling issues with respect to the complexity of the specifications and the potential occurrence of non-differentiable terms, classical optimization methods often solve STL-based problems inefficiently. Smoothing and approximation techniques can alleviate these issues but require changing the optimization problem. This paper proposes a novel sampling-based method based on model predictive path integral control to solve optimal control problems with STL cost functions. We demonstrate the effectiveness of our method on benchmark motion planning problems and compare its performance with state-of-the-art methods. The results show that our method efficiently solves optimal control problems with STL costs.
	\end{abstract}

	\copyrightnotice  

	\section{Introduction}\label{sec:Introduction}
\begin{figure}[!tb]
    \centering
    \resizebox{0.75\columnwidth}{!}{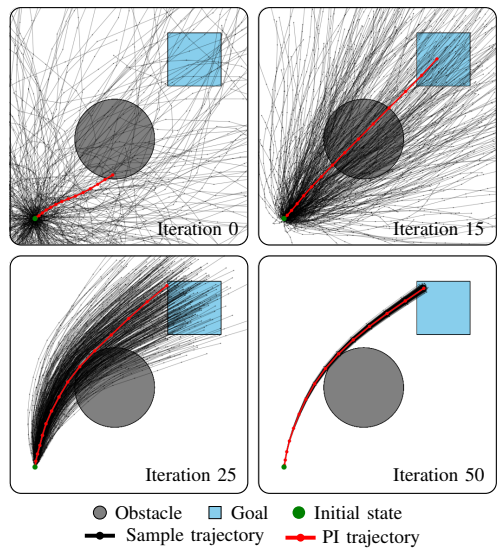}
    \vspace*{-3mm}
    \caption{Example of our PI-based solution for solving an optimal control problem with STL costs. The objective is to reach the blue box \textit{eventually} while ensuring that the gray circular obstacle is avoided \textit{at all times}. Four intermediate solutions are presented.}
    \label{fig:Initial_Example}
    \vspace*{-6mm}
\end{figure}

Rich specifications are essential for delineating the desired behavior of dynamic systems~\cite{Mehdipour2023}. Formal specification languages, such as signal temporal logic (STL), provide a precise yet interpretable way of describing the desired behavior, making them particularly attractive in robotics and autonomous driving~\cite{Maierhofer2020}. Due to the inherent robustness measure, the level of satisfaction of an STL specification can straightforwardly be considered in optimal control problems. However, solving such problems often presents challenges due to a) scalability issues when STL specifications are deeply nested or contain long time intervals and b) potentially non-differentiable terms in the STL robustness measure. We address these issues by introducing a sequential solving method based on the path integral (PI) control framework~\cite{Williams2018}. Our approach converges quickly by sampling in the input space and gradually reducing the sampling variance, as illustrated in Fig.~\ref{fig:Initial_Example}.

\subsection{Related Work}\label{sec:sota}
Subsequently, we review related work on optimization with STL and model predictive path integral control. 

\paragraph{Optimization with STL}\label{sec:STLMotionPlanning}

STL is well suited for formalizing spatio-temporal requirements in motion planning since it provides a precise, mathematical way of defining the desired behavior but remains interpretable by \mbox{humans~\cite{Hekmatnejad2019, Maierhofer2020}}. STL provides a robustness semantic, quantifying the degree to which a formula is satisfied or violated~\cite{Bartocci2018}. A respective robustness function can be automatically derived, eliminating the need for manual design. STL specifications are usually integrated into optimization problems by including the robustness in the constraints or considering it in the objective as a penalty. However, solving these problems is challenging, e.g., due to the non-differentiable terms in the robustness function~\cite{Kurtz2022}.

STL robustness is typically optimized using classical optimization techniques~\cite{Belta2019}. Mixed-integer encodings of the robustness function are mainly utilized (e.g., see~\cite{Sadraddini2019, Sahin2020,  Cardona2023}), but gradient-based methods~\cite{Pant2017, Mehdipour2019, Mao2022} are also often applied. Recently, control barrier function approaches~\cite{Lindemann2019, Lindemann2020, Charitidou2021, Xiao2021} and funnel approaches~\cite{Lindemann2017, Lindemann2021} have also gained popularity. However, several problems arise with the aforementioned methods. Especially for mixed-integer encodings, scalability with respect to the time horizon and the nesting of the STL formula is a significant issue, as the number of constraints increases exponentially~\cite{Kurtz2022}.

For gradient-based and control barrier function methods, the non-differentiable terms of the robustness function present another big challenge because of potentially vanishing gradients~\cite{Leung2021}. Thus, these approaches often require significantly modifying the original optimization problem, e.g., by restricting to a fragment of STL~\cite{Lindemann2021}, limiting to convex predicates~\cite{Kurtz2022, Halder2023}, or using a modified smoothed robustness definition (see~\cite{Welikala2023} for a comparison). STL is also used in learning-based optimization methods~\cite{Leung2021}, which are also affected by the same problems. Sampling-based~\cite{Karlsson2020, Linard2023} and evolutionary methods~\cite{Alsalehi2022} are barely used for optimization with STL constraints, possibly because they only provide asymptotic optimality and may suffer from slow convergence rates.

\paragraph{MPPI Control}\label{sec:MPPIControl}
 Model predictive path integral (MPPI) control~\cite{Williams2018} is a sampling-based approach that is suitable for trajectory planning and model predictive control. In contrast to most first- and second-order optimization methods employed for model predictive control~\cite{Nocedal2006, Rawlings2017}, MPPI  can be straightforwardly applied to systems with  non-differentiable costs and dynamics, is easy to implement, and can be inherently executed in parallel~\cite{Williams2017}.

However, similar to other sampling-based methods, like reward-weighted regression~\cite{Peters2007}, or the covariance matrix adaptation evolution strategy (CMA-ES) algorithm~\cite{Hansen2006}, MPPI can struggle in high-dimensional problems. To overcome this, recent extensions focus on improving the sampling distribution, e.g., by improving the convergence employing gradient descent updates~\cite{Okada2018}, adaptive importance sampling~\cite{Kappen2016}, learning input distributions to obtain low-cost samples~\cite{Kusumoto2019}, covariance steering~\cite{Balci2022}, handcrafted skills~\cite{Homburger2022b}, or ancillary controllers~\cite{Trevisan2024}. MPPI can be improved by combining it with iterative linear-quadratic Gaussian~\cite{Gandhi2021} or differential dynamic programming~\cite{Lefebvre2019}.

While MPPI was applied initially to control unmanned aerial vehicles~\cite{Gomez2016} and an intelligent race car~\cite{Williams2016, Williams2017, Williams2018}, recently, it was also intensively used by the robotics community to control systems, such as autonomous surface vessels~\cite{Homburger2022, Streichenberg2023}, autonomous underwater vehicles~\cite{Nicolay2023}, robotic manipulators~\cite{Hou2022, Yamamoto2022}, four-legged walking robots\cite{Carius2022}, and lab experiments like the Furuta pendulum~\cite{Homburger2022a, Trevisan2024}. A fast GPU implementation called \texttt{MPPI-Generic} is presented in \cite{Vlahov2024}. For a comprehensive overview of recent methodological contributions and applications of MPPI, the interested reader is referred to~\cite{Kazim2024}.

\subsection{Contributions}\label{sec:Contributions}

This paper presents a novel PI-based approach for solving deterministic, discrete-time optimal control problems with finite horizon and STL costs. In detail, our contributions are:

\begin{itemize}
    \item enabling a simple task formalization and planning process by combining STL and MPPI control; 
    \item providing an algorithm to determine the solution for deterministic discrete-time finite-horizon nonlinear optimal control problems with arbitrary STL costs; and
    \item comparing our approach numerically with state-of-the-art methods for trajectory planning with STL costs.
\end{itemize}
The remainder of this paper is organized as follows: we formally define our system, classical MPPI, and STL  in Sec.~\ref{sec:Preliminaries}, introduce the problem statement in Sec.~\ref{sec:ProblemStatement}, and provide our solution in Sec.~\ref{sec:Solution}. Numerical experiments are presented in Sec.~\ref{sec:NumericalExperiments}, and we conclude in Sec.~\ref{sec:Conclusions}. 

	\section{Preliminaries}\label{sec:Preliminaries}

\subsection{System Definition and Projection Operator}
Let $k \in \mathbb{N}_0$ be a discrete time step and $t_k:= k\Delta t$ the corresponding continuous time, with $\Delta t \in \mathbb{R}_{> 0}$. Without loss of generality, the initial time step is $k_0 = 0$, and the final time step is $K \in \mathbb{N}_0$. Further, let ${\mathcal{K} := [0, K] \subseteq \mathbb{N}_0}$ be a discrete time interval. We consider a deterministic nonlinear discrete-time system
\begin{equation}\label{eq:SystemDefinition}
    x_{k+1} = f(x_k, u_k),
\end{equation}
where $x_k \in \mathbb{R}^{n_x}$ and $ u_k \in  \mathbb{R}^{n_u}$ represent the state and input, respectively, with $k \in \mathcal{K}$. A solution of (\ref{eq:SystemDefinition}) for an initial state $x_0 \in \mathbb{R}^{n_x}$ and an input trajectory $\bu=\left[u_0,u_1,\dots,u_{K-1}\right]$ at time step $k\in\mathcal K$ is denoted by $\chi (x_0, \bu, k)$ and subsequently, we use $ x_k$ as its shorthand notation. The state trajectory is denoted by  $\bx=\left[x_0,x_1,\dots,x_{K}\right]$.

Further, we introduce the projection operator $\proj_{\square}: \Theta \rightarrow \mathbb{R}$ which maps a vector $\varrho \in \Theta$ to its element specified by $\square \, $, and $\boldsymbol{0}_{n_0}$ is a vector of zeros of length $n_0 \in \mathbb{N}_{>0}$.

\subsection{Signal Temporal Logic}\label{sec:SignalTemporalLogic}
In this work, we use the future-time fragment of STL; however, our approach is also analogously applicable to the past-time fragment. The syntax of the future-time fragment of STL over a state trajectory $\bx$ is defined by the following grammar \mbox{\cite[Sec. 2.1]{Bartocci2018}}:
\begin{equation*}
    \varphi:= \mu \; \vert\; \lnot \varphi\; |\; \varphi_1 \land \varphi_2 \;|\; \varphi_1 \mathbf{U}_{I} \varphi_2,
\end{equation*}
where $\mu$ is a predicate of the form ${\mu := b( \bx, k) \geq 0}$, with $b: \mathbb{R}^{n_x \times (K+1)} \times \mathbb{N}_0 \rightarrow \mathbb{R}$. The expressions $\varphi, \varphi_1$, and $\varphi_2$ are STL formulas, and $\lnot$ and $\land$ are negation and conjunction, respectively. Disjunction can be expressed as $\varphi_1 \lor \varphi_2 := \lnot(\lnot \varphi_1 \land \lnot \varphi_2)$. The temporal operator $\mathbf{U}_{I}$ specifies that $\varphi_1$ holds \textit{until} $\varphi_2$ holds in the time interval $[k_{\min}, k_{\max}], \text{where } k_{\min}, k_{\max} \in \mathbb{N}_0 \text{ and } k_{\max} \geq k_{\min}$. Further temporal operators can be derived, such as ${\mathbf{F}_{I}\varphi := \mathrm{True}\mathbf{U}_{I} \varphi}$ (\textit{eventually}) and ${\mathbf{G}_{I}\varphi := \lnot \mathbf{F}_{I}\lnot\varphi}$ (\textit{globally}). The satisfaction of an STL formula $\varphi$ by a trajectory $ \bx$ at time step $k$ is denoted as $(\bx, k) \models \varphi$.

Subsequently, we present the definition of \textit{space robustness} for the operators used in this paper. The space robustness $\rho^{\varphi}( \bx, k) \in \mathbb{R}$ of an STL formula $\varphi$ provides a measure expressing its degree of compliance or violation for a finite trajectory $ \bx$ at time step $k \in \mathcal{K}$. The robustness $\rho^{\varphi}( \bx, k)$ is positive iff $ \bx \models \varphi$. The space robustness is formally defined as \cite[Sec. 2.2]{Bartocci2018}:
\begin{align*}
    \rho^{\mathrm{True}} ( \bx, k)                       &:= \infty,\\
    \rho^{b( \bx, k) \geq 0} ( \bx, k)                   &:= b( \bx, k),\\
    \rho^{\lnot\varphi} ( \bx, k)                        &:= -\rho^{\varphi} ( \bx, k),\\
    \rho^{\varphi_1 \land \varphi_2} ( \bx, k)           &:= \min\big(\rho^{\varphi_1} ( \bx, k), \rho^{\varphi_2} ( \bx, k)\big),\\
    \rho^{\mathbf{G}_{I} \varphi} ( \bx, k)              &:= \min_{k^\prime \in [k+k_{\min}, k+k_{\max}] \cap \mathcal{K}}\big(\rho^{\varphi} ( \bx, k^\prime)\big),\\
    \rho^{\mathbf{F}_{I} \varphi} ( \bx, k)              &:= \max_{k^\prime \in [k+k_{\min}, k+k_{\max}] \cap \mathcal{K}}\big(\rho^{\varphi} ( \bx, k^\prime)\big).
\end{align*}

\subsection{Classical MPPI}\label{sec:MPPI}
Let us recall the basic ideas of classical MPPI control derived from the information-theoretic framework \cite{Williams2018}.
MPPI considers the stochastic open-loop optimal control problem
\begin{equation}\label{eq_SOCP}
    \bu^*=\argmin_{\bu} \mathbb{E}_{\mathbb{Q}_{\bu}}\left[ \sum_{k=0}^{K-1} L(  x_k,u_k)  + E(  x_K) \right],
\end{equation}
where a disturbed state trajectory is given by $\bx(x_0, \bv, k)\; \text{for}\; k \in \mathcal{K}$, originating from the disturbed input sequence $\bv=[v_0, v_1, \dots,v_{K-1}]$, with $v_k \sim \mathcal{N}(u_k,\Sigma)$. The covariance matrix is denoted by $\Sigma\in\mathbb{R}^{n_u\times n_u}$, $\mathbb{Q}_{\bu}$ is the Gaussian distribution of the sequence $\bv$ with mean~$\bu$, the corresponding probability density function is $q(\bv|\bu)$, and  $\mathbb{E}_{\mathbb{Q}_{\bu}}[\cdot]$ denotes the expectation under distribution $\mathbb{Q}_{\bu}$. Further, $E:\mathbb{R}^{n_x}\rightarrow\mathbb{R}_{>0}$ is the terminal cost function and the stage cost function $L(x_k,u_k)$ is chosen as
\begin{equation}\label{eq_cost}
    L(x_k,u_k) :=c(  x_k)+\frac{1}{2} u_k^\top R u_k,
\end{equation} where $c:\mathbb{R}^{n_x}\rightarrow\mathbb{R}_{\geq 0}$ and the system input is penalized quadratically with $R:=\lambda\Sigma^{-1}$, where $\lambda\in\mathbb{R}_{>0}$ is a tuning parameter called the \textit{inverse temperature} \cite[Sec. 2]{Williams2018}. 
Based on these assumptions, according to~\cite[Sec. 3.A]{Williams2018},
the optimal control sequence $\bu^*$ that solves \eqref{eq_SOCP} is approximated by the so-called \textit{information-theoretic} optimum $\tilde \bu^*$:
\begin{equation}\label{eq_KL}
    \bu^* \approx  \argmin_{\bu} \mathbb{E}_{\mathbb{Q}^*}\left[ \log \left( \frac{q^*(\bv)}{q(\bv|\bu)}\right) \right]=: \tilde \bu^*, 
\end{equation}
with the optimal distribution defined by its probability density function 
\begin{equation}\label{eq_opt_dis}
    q^*(\bv):=\frac{1}{\eta} \exp \left(-\frac{1}{\lambda} S(\bv) \right) q(\bv|\boldsymbol{0}_{Kn_u}),
\end{equation}
where $\eta\in\mathbb{R}_{>0}$ is a normalization constant, and the path costs are
\begin{equation*}
    S(\bv) := E(  x_K)+\sum_{k=0}^{K-1} c(x_k).
\end{equation*} 
The equation of the information-theoretic optimum in \eqref{eq_KL} can be simplified by disregarding constant terms and using the expected value of a random variable from a distribution $\mathbb{Q}_{\hat{u}}$, where $\hat \bu$ is a proposed control sequence, along with a correction term $w(\bv, \hat{\bu}) \in \mathbb{R}_{>0}$ (see~\cite[Sec. 3]{Williams2018} for details). This is then given by
\begin{align}
   \tilde u^*_k&= \mathbb{E}_{\mathbb{Q}^\star}\left[  v_k \right]=\mathbb{E}_{\mathbb{Q}_{\hat \bu}}\left[ w(\bv,\hat \bu) v_k \right]  
         \label{eq_opt_control}
\end{align}
for $k \in \mathcal{K}$. 
It is common to estimate the so-called optimal \textit{information-theoretic} control law \eqref{eq_opt_control} by Monte Carlo sampling of trajectories.

	\section{Problem Statement}\label{sec:ProblemStatement}
Let us first define a robustness function $\rob:\mathbb{R}^{(K+1)n_x}\rightarrow \mathbb{R}$. Two prevalent choices are 1) to either maximize the satisfaction of $\varphi$ or 2) to impose no costs if $\varphi$ is satisfied and to minimize its violation otherwise:
\begin{subequations}
    \begin{align}
        \rob^{\varphi}( \bx) &:= -\rho^{\varphi}(  \bx, 0) \label{eq:rob_cost_fct_max}, \\
        \rob^{\varphi}( \bx) &:= -\min(0, \rho^{\varphi}( \bx, 0)). \label{eq:rob_cost_fct_viol}
    \end{align}
\end{subequations}
We solve the nonlinear program typically arising from the direct single shooting discretization of an optimal control problem, with an additional STL cost term $\rob$ as
\begin{equation}\label{eq_STL_NLP}
    \min_{u_0,\dots,u_{K-1}}  \gamma \rob^{\varphi}( \bx) +\sum_{k=0}^{K-1} L(  x_k,u_k)  + E(  x_K),
\end{equation}
where $\gamma \in \mathbb{R}_{>0}$ is a weight. We call \eqref{eq_STL_NLP} an STL-OCP. The challenge in solving this problem stems from the non-differentiable and  non-convex function $\rob$. However, many state-of-the-art methods (e.g., see ~\cite{Nocedal2006, Rawlings2017}) require differentiable and convex components for a good performance.  
The feasibility of \eqref{eq_STL_NLP} is always guaranteed since it is an unconstrained problem.
Subsequently, we present our solution method to solve problem~\eqref{eq_STL_NLP}.

	\section{Solution}\label{sec:Solution}

As presented in Sec.~\ref{sec:sota}, MPPI is a prevalent sampling-based method to solve optimal control problems, which is straightforward to implement, can be executed in parallel, and can handle non-differentiable costs and dynamics. Motivated by these properties, we develop a PI-based solution method for the STL-OCP~\eqref{eq_STL_NLP}. 

Considering the STL-OCP \eqref{eq_STL_NLP} compared to the standard MPPI optimal control problem \eqref{eq_SOCP}, two differences occur: the additional STL cost term $\rob$, which depends on the entire state trajectory, and the deterministic problem setting. We thus show subsequently how the STL-OCP \eqref{eq_STL_NLP} can be transformed into the required form, how the exact solution for our deterministic problem can be determined using the stochastic PI method, and finally, we present our solution algorithm.

\subsection{Problem Transformation}

The STL-OCP \eqref{eq_STL_NLP} cannot be directly solved with MPPI since the standard discrete-time Bolza form~\cite[Sec. 1]{Rockafellar1983} is required. Therefore, we introduce an augmented state $\tilde{x}_k \in \mathbb{R}^{(K+1)n_x}$ and a corresponding augmented system dynamics
$\tilde{f}:\mathbb{R}^{(K+1)n_x}\times \mathbb{R}^{n_u}\rightarrow \mathbb{R}^{(K+1)n_x}$, defined as
\begin{equation}\label{eq_aug_state}
    \begin{aligned}
        \tilde{x}_k &:= \renewcommand{\arraystretch}{0.9} \begin{bmatrix}
               x_k \\
               x_{k-1} \\
             \vdots \\
                x_0 \\
             \boldsymbol{0}_{(K-k)n_x}
        \end{bmatrix},\
        \tilde{f}(\tilde{x}_k, u_k) := \renewcommand{\arraystretch}{0.7} \begin{bmatrix}
             f(  x_k, u_k) \\
               x_k \\
               x_{k-1} \\
               \vdots \\
               x_0 \\
            \boldsymbol{0}_{(K-k-1)n_x} 
        \end{bmatrix}.
    \end{aligned}
\end{equation}
The augmented state $\tilde{x}_k$ concatenates the original states for all time steps up to the current time step $k$. The zero vectors assure constant state dimensions. The cost terms are defined as
\vspace{-0.25cm}
\begin{align*}
    \tilde{L}(\tilde{x}_k, u_k) &:= L(  x_k, u_k),\\
    \tilde{c}(\tilde{x}_k)&:=c(  x_k),\\
    \tilde{E}(\tilde{x}_K) &:= E(  x_K) + \gamma \rob^{\varphi}(\tilde{x}_K),
\end{align*}
where $\tilde{x}_K \equiv \bx$ is exploited.
Now, \eqref{eq_STL_NLP} can be expressed in standard discrete-time Bolza form as 
\begin{equation}\label{eq_STL_lifted_NLP}
    \begin{aligned}
        \min_{u_0, \dots, u_{K-1}} &\sum_{k=0}^{K-1} \tilde{L}(\tilde{x}_k, u_k) + \tilde{E}(\tilde{x}_K) \\
    \end{aligned}
\end{equation}
based on the state augmentation \eqref{eq_aug_state}.
The new representation~\eqref{eq_STL_lifted_NLP} has the desired form, but it comes with the drawback of a high-dimensional state space. As we will see later, this does not pose a problem for our solution approach since trajectories are only sampled in the input space. 


\subsection{Exact Solution for Deterministic Problems with MPPI}
While the standard MPPI algorithm in~\cite{Williams2018} generally provides a biased solution to the open-loop stochastic optimal control problem \eqref{eq_KL} \cite[Sec. 3.B]{Homburger2025}, we now present a method to obtain the unbiased solution for the deterministic STL-OCP \eqref{eq_STL_lifted_NLP} by gradually reducing the uncertainty $\Sigma$ and the inverse temperature $\lambda$.

Assuming there is a unique global minimizer 
and considering the optimal distribution \eqref{eq_opt_dis}, where we substitute the probability density function $q(\bv|\boldsymbol{0}_{Kn_u})$ by its explicit expression, we obtain
\begin{align*}
    q^*(\bv)&=\frac{1}{\eta} \exp \left( -\frac{1}{\lambda} S(\bv) \right) \frac{1}{Z^{K}} \prod_{k=0}^{K-1}\exp \left(  -\frac{1}{2} v_k^\top \Sigma^{-1} v_k \right)\\
    &=\frac{1}{\tilde \eta} \exp \left( -\frac{1}{\lambda}\left[ S(\bv)  +\frac{1}{2} \sum_{k=0}^{K-1} v_k^\top \lambda\Sigma^{-1} v_k\right] \right),
\end{align*}
where $Z=\left[(2\pi)^{n_u}\det (\Sigma)\right]^{\frac{1}{2}}$ is the normalization constant for the multivariate normal distribution. The second line originates from the definition of the assembled normalization constant $\tilde \eta = \eta Z^K$ and by applying the standard laws of exponents. Substituting $\lambda$ by $\beta\lambda$ and $\Sigma$ by $\beta\Sigma$, where $\beta\in (0,1)$ is a scaling factor, yields
\begin{equation}
     q^*(\bv,\beta)=\frac{\exp \left( -\frac{1}{\beta\lambda}  \left[ S(\bv)  +\frac{1}{2} \sum_{k=0}^{K-1} v_k^\top \lambda\Sigma^{-1} v_k\right] \right)}{\tilde \eta(\beta)} .
\end{equation}
Note that the expression within the square brackets reflects the objective of \eqref{eq_STL_lifted_NLP}. By definition, this expression is minimal for $\bv\equiv\bu^*$. By shrinking $\beta$, the effects of the suboptimal trajectories on $q^*(\bv,\beta)$ decrease, resulting in
\begin{equation}\label{eq_opt_pdf}
  \bu^*= \lim_{\beta\rightarrow 0} \mathbb{E}_{\mathbb{Q}^*(\beta)} \left[ \bv \right] = \lim_{\beta\rightarrow 0} \tilde \bu(\beta) . 
\end{equation}
We refer the reader to~\cite{Homburger2025} for a comprehensive analysis of the convergence properties.
Subsequently, we present an algorithm to numerically compute $\bu^*$ based on a progressive reduction of $\beta$. 


\subsection{Algorithm}

Alg.~\ref{alg:DeterministicSTLMPPISolver} solves problem \eqref{eq_STL_NLP}. In contrast to the standard MPPI algorithm~\cite{Williams2017}, the covariance $\Sigma$ and the inverse temperature $\lambda$  gradually decrease within $J \in \mathbb{N}_{> 0}$ iterations by multiplying with the shrinking factor $\nu \in (0,1)$ (see Lines~\ref{eq:Det_MPPI_Shrink_Loop_Start},~\ref{eq:Det_MPPI_Shrink_Lambda}~and~\ref{eq:Det_MPPI_Shrink_Sigma}). This approximates the weak limit of $\beta\rightarrow 0$ in \eqref{eq_opt_pdf}. In each iteration, $M \in \mathbb{N}_{> 0}$ trajectories are simulated, and their corresponding costs are evaluated (see Lines~\ref{eq:Det_MPPI_MPPI_Loop_Start} to~\ref{Det_MPPI_MPPI_Loop_End}). The correction term $\lambda {\epsilon^m_k}^\top \Sigma^{-1} \hat u_{k}$ in Line~\ref{eq:cost_with_importance_weighting} corresponds to the importance weighting of MPPI (cf.~\eqref{eq_opt_control}). The solution trajectory for one iteration is determined using the weighted sum of the sampled input trajectories (see Lines~\ref{eq:Det_MPPI_MPPI_Weighting_Start} to~\ref{eq:Det_MPPI_MPPI_Weighting_End}). The output of the algorithm is the solution of the optimal control problem for a reduced covariance $\Sigma\nu^{J}\approx 0$.



Alg.~\ref{alg:DeterministicSTLMPPISolver} is characterized by gradually reducing the input uncertainty, allowing for a good tradeoff between an intense exploration of the solution space and converging fast towards a local minimum. The algorithm scales linearly with the product of the number of shrinking iterations $J$, the number of trajectory samples $M$, and the problem horizon $K$, where $M$ is typically large. By using a parallel implementation of the loop in Lines~\ref{eq:Det_MPPI_Shrink_Loop_Start} to~\ref{Det_MPPI_MPPI_Loop_End}, we can alleviate the runtime dependency on $M$, which usually dominates the computational cost of Alg.~\ref{alg:DeterministicSTLMPPISolver}. The algorithm can be terminated early for real-time applications with limited runtime since intermediate solutions are computed.

While the standard MPPI algorithm~\cite{Williams2018} or the deterministic \texttt{MPPI-Generic} algorithm~\cite{Vlahov2024} provide suboptimal solutions, Alg.~\ref{alg:DeterministicSTLMPPISolver} solves the deterministic STL-OCP~\eqref{eq_STL_NLP} up to an arbitrary accuracy \cite[Thm. 1]{Homburger2025}.
The convergence of Alg.~\ref{alg:DeterministicSTLMPPISolver} as ${M\rightarrow \infty}$ and ${J\rightarrow \infty}$ is proven in \cite[Thm. 1]{Homburger2025}.
Contrary to the standard MPPI, $\Sigma$ and $\lambda$ are hyper-parameters without physical meaning. In contrast to other Monte-Carlo methods like, e.g., reward-weighted regression~\cite{Peters2007} and CMA-ES~\cite{Hansen2006}, our approach only adapts the mean of the sample distribution, whereas the exponential shrinking of the covariance is predefined. 

\begin{algorithm}[!b]
    \caption{Deterministic STL-OCP Path Integral Solver}  \label{alg:DeterministicSTLMPPISolver}
    {\small
	\begin{algorithmic}[1]
        \setlength{\itemsep}{1pt}  
        \Require Initial system state $x_0$, initial control sequence $\hat \bu$, number $J$ of iterations, number $M$ of trajectory samples, shrinking factor $\nu$, initial covariance $\Sigma$, initial inverse temperature $\lambda$
        \Ensure  Optimal input trajectory $\bu^*$, optimal state trajectory $\bx^*$
        \For {$j \in \{1, 2, \dots, J\}$} \label{eq:Det_MPPI_Shrink_Loop_Start}
            \For {$m\in \{1, 2, \dots, M\}$ }  \Comment{parallelizable} \label{eq:Det_MPPI_MPPI_Loop_Start}
                \State $\tilde{x}^m_{0} \leftarrow x_0$
                \State $S^m\gets 0$
                \For {$k \in \{0, 1, \dots, K-1\}$}
                    \State $\epsilon_k^m \sim \mathcal{N}(0,\Sigma) $
                    \State  $\tilde{x}^m_{k+1} \leftarrow \tilde{f}(\tilde{x}^m_{k}, \hat{u}_{k} + {\epsilon}_{k}^m)$
                    \State $S^m \gets S^m + \tilde c(\tilde{x}^m_{k}) + \lambda {\epsilon^m_k}^\top \Sigma^{-1} \hat u_{k} $ \label{eq:cost_with_importance_weighting}
                \EndFor
                \State $S^m \gets S^m + \tilde{E}(\tilde{x}^m_{K})$
            \EndFor \label{Det_MPPI_MPPI_Loop_End}
            \State $\psi \gets \min_{m \in \{1, 2, \dots , M\}} S^m$ \label{eq:Det_MPPI_MPPI_Weighting_Start}
            \State $\eta \gets \sum_{m=1}^{M} \exp (-\frac{1}{\lambda}(S^m - \psi))$
            \For {$m \in \{1, 2, \dots, M\}$}
                \State $\omega^m \gets \frac{1}{\eta} \exp (-\frac{1}{\lambda}(S^m - \psi))$
            \EndFor
            \For {$k\in\{0, 1, \dots, K-1\}$}
                \State $\hat u_k \gets \hat u_k + \sum_{m=1}^{M} \omega^m  \epsilon_k^m $
            \EndFor \label{eq:Det_MPPI_MPPI_Weighting_End}
            \State $\lambda \gets \nu\lambda$ \label{eq:Det_MPPI_Shrink_Lambda}
            \State $\Sigma \gets \nu \Sigma$\label{eq:Det_MPPI_Shrink_Sigma}
        \EndFor
    \State $\bu^* \gets \hat \bu, x^*_0 \gets x_0$ 
    \For {$k \in \{0, 1, \dots, K-1\}$}
        \State $x^*_{k+1} \gets f(x^*_k, u_k^*)$
    \EndFor
    \State \Return $\bu^*, \bx^*$
    \end{algorithmic}}
\end{algorithm}

	\section{Numerical Experiments}\label{sec:NumericalExperiments}
In this section, STL motion planning problems are solved numerically to evaluate the performance of the PI approach and to compare it with state-of-the-art solvers. In particular, we compare the following solvers:

\begin{itemize}
    \item \textbf{MIP}: Mixed-integer encoding of robustness, typically used for linear systems~\cite{Belta2019}, but also applicable to nonlinear systems, solved with \texttt{Gurobi}~\cite{GurobiOptimization2024};
    \item \textbf{GRAD}: Gradient-based solver using the \texttt{SLSQP} method from \texttt{SciPy}~\cite{Virtanen2020};
    \item \textbf{SGRAD}: Smooth approximation of robustness~\cite{Gilpin2021}; solved with \texttt{SLSQP} from \texttt{SciPy}~\cite{Virtanen2020};
    \item \textbf{CMA-ES}: Adaptive evolutionary solver from~\cite{Blank2020};
    \item \textbf{PI}: The approach of this paper; see Alg.~\ref{alg:DeterministicSTLMPPISolver}. 
\end{itemize}
Our algorithm is implemented in C++, and the experiment setup is based on the STL-benchmark tool \texttt{stlpy}~\cite{Kurtz2022}.  
For comparability, we extensively optimize the hyperparameters of each solver with \texttt{Optuna}~\cite{Akiba2019}, by minimizing the objective $C^* + \vartheta t_\mathrm{c}^*$, where $C^*$ is the optimized cost value of \eqref{eq_STL_NLP}, $t_\mathrm{c}^*$ is the convergence time, and $\vartheta$ is a tuning parameter, with $\vartheta_{\text{Prob. I}}= 0.1$, $\vartheta_{\text{Prob. II}}= 1.0$, and $\vartheta_{\text{Prob. III}}= 0.1$. We terminate after $\texttt{n\_trials} = 1000$. The hyperparameters of the \textbf{MIP} solver are optimized using the tuning tool of \texttt{Gurobi}, specifically designed for this purpose, and no improvements are detected compared to the default values. The optimized hyperparameters are listed in Tab.~\ref{tab:Parametization}.
\begin{table}[!t]
\vspace{3mm}
    \caption{Optimized hyperparameters of the used solvers.}
    \vspace{-3mm}
    \renewcommand{\arraystretch}{1.0}
    \centering
    \begin{adjustbox}{max width=1.0\columnwidth}
        \begin{tabular}{cllll}
            \toprule
            \textbf{Solver}                 & \textbf{Parameters}    & \textbf{Pr. I}   & \textbf{Pr. II}  & \textbf{Pr. III}  \\
            \midrule
            \textbf{MIP}                    & all parameters                   &\multicolumn{3}{c}{- defaults -}\\
            \midrule
            \multirow{3}{*}{\textbf{GRAD}}  & \texttt{maxiter} & $100$                   & $100$                  & $1 \mathrm{e}{5}$\\
                                            & \texttt{ftol}    & $7.0 \mathrm{e}{-5}$   & $2.2 \mathrm{e}{-7}$  & $1 \mathrm{e}{-6}$\\
                                            & \texttt{eps}     & $4.7 \mathrm{e}{-5}$   & $1.6 \mathrm{e}{-7}$  & $1.49 \mathrm{e}{-8}$\\
            \midrule                      
            \multirow{3}{*}{\textbf{SGRAD}} & \texttt{maxiter}   & $100$                & $30$                  & $1 \mathrm{e}{5}$ \\
                                            & \texttt{ftol}      & $3.4 \mathrm{e}{-6}$ & $6.7 \mathrm{e}{-7}$  & $1 \mathrm{e}{-6}$ \\
                                            & \texttt{eps}       & $1.5 \mathrm{e}{-5}$ & $4.8 \mathrm{e}{-7}$    & $1.49 \mathrm{e}{-8}$ \\
                                            & \texttt{k\_1, k\_2}& $186$                & $490$                 & $400$ \\
            \midrule
            \multirow{3}{*}{\textbf{CMA-ES}} & \texttt{sigma}      & $0.03$             & $0.037$                & $0.022$ \\
                                             & \texttt{maxfevals}  & $4060$            & $9300$               & $1 \mathrm{e}{6}$ \\
                                             & \makecell[l]{\texttt{noise\_change}\_\\\texttt{sigma\_exponent}} & $0.887$  & $0.555$ & $0.644$ \\
                                             & \texttt{pop\_size}  & $17$               & $13$                   & $35$ \\
                                             & \texttt{n\_iter}    & $2140$               & $773$               & $11240$ \\
            \midrule
            \multirow{3}{*}{\textbf{PI}}    & $J$       & $19$      & $75$    & $40$ \\
                                            & $M$       & $955$    & $1140$   & $81650$ \\
                                            & $\Sigma$  & $5.6$   & $\left[\begin{smallmatrix} 3.4 & 0.0 \\ 0.0 & 3.4 \end{smallmatrix}\right]$   & $\left[\begin{smallmatrix} 0.002 & 0.0 \\ 0.0 & 0.002 \end{smallmatrix}\right]$ \\ 
                                            & $\lambda$ & $11.2$     & $60.8$   & $0.2$ \\ 
                                            & $\nu$     & $0.3$     & $0.8$   & $0.8$ \\ 
            \bottomrule
            \end{tabular}
        \label{tab:Parametization}
    \end{adjustbox}
    \vspace{-5mm}
\end{table}

For problem~I and problem~III, \eqref{eq:rob_cost_fct_viol} is the robustness cost function and \eqref{eq:rob_cost_fct_max} is selected for problem~II. For all experiments, we use $c(x_k) := 0$. The experiments are executed on an \texttt{AMD Ryzen\textsuperscript{TM} 5 PRO 7540U CPU}. The code can be accessed at \href{https://github.com/TUMcps/STL-PI-Planner}{\mbox{https://github.com/TUMcps/STL-PI-Planner}}.

\begin{table}[!tb]
    \vspace{2.8mm}
    \caption{Optimal costs, robustness, and convergence times.} 
    \vspace{-3mm}
    \renewcommand{\arraystretch}{1.0}
    \centering
    \begin{adjustbox}{max width=1.0\columnwidth}
        \begin{tabular}{clllllll}
            \toprule
             \textbf{Pr.} &  & \textbf{MIP} & \textbf{GRAD} & \textbf{SGRAD} & \textbf{CMA-ES} & \textbf{PI} \\
            \midrule
            \multirow{3}{*}{\textbf{I}}   & ${C}^*$ & $-3.0$ & $-0.25$& $-2.85$& $-2.99$& $-2.98$ \\
                                 & $\rho^{\varphi_1}$   & $0.0$ & $-2.5$ & $0.0$ & $0.0$ & $0.0$ \\
                                & $t_\mathrm{c}^*$   & \SI{0.008}{\second} & \SI{0.039}{\second} & \SI{0.104}{\second} & \SI{0.397}{\second} & \SI{0.024}{\second} \\
            \midrule
            \multirow{3}{*}{\textbf{II}}   & ${C}^*$ & $13.32$ & $13.33$ & $13.58$ & $13.4$ & $13.48$ \\
                                 & $\rho^{\varphi_2}$   & $0.2$ & $0.2$ & $0.015$ & $0.2$ & $0.17$ \\
                                & $t_\mathrm{c}^*$ & \SI{1.16}{\second} & \SI{1.27}{\second} & \SI{0.34}{\second} & \SI{1.62}{\second} & \SI{0.12}{\second}\\
            \bottomrule
        \end{tabular}
        \label{tab:PerformanceResults}
    \end{adjustbox}
    \vspace{-5mm}
\end{table}
\subsection{Problem I: Scalar Integrator}

Let us consider the following linear scalar discrete-time integrator with the dynamics $  x_{k+1} = x_k + u_k$.
We aim to maximize $x_K$ at the end of the planning time horizon ${K=10}$. Additionally, the state shall be below $1$ for at least two not necessarily consecutive time steps. This problem could be formalized as a mixed-integer linear program; however, specifying the required constraints is cumbersome, and solving mixed-integer problems is hard. 

Hence, we formalize an STL formula and consider it in the cost function. We define ${E(x_K):= -  x_K}$, and the STL formula is $ \varphi_1 := F_{[0,K]} (\mu_{\text{gate}} \land F_{[1,K]} (\mu_{\text{gate}}))$, with $\mu_{\text{gate}} := 1 - x_k \geq 0$.
The corresponding robustness function can be directly derived from the grammar shown in Sec.~\ref{sec:SignalTemporalLogic} as:
\begin{align*}
   \rho^{\varphi_1}(\bx, k) =& \max_{k' \in [k, k+K] \cap \mathcal{K}} \Big( \min \big(1 -  x_{k'}, \\
    & \max_{k'' \in [k' + 1, k' + K] \cap \mathcal{K}} ( 1 -   x_{k''} ) \big) \Big).
\end{align*}
This non-differentiable function is costly to evaluate and hard to maximize. 

\begin{figure}[!b]
    \vspace*{-5mm}
    \centering
    \resizebox{0.98\linewidth}{!}{\begin{tikzpicture}
    \definecolor{blue}{RGB}{2,2,255} 
    \definecolor{green}{RGB}{0,128,0} 
    \definecolor{orange}{RGB}{255,166,0}     
    \definecolor{purple}{RGB}{130,0,130} 
    \definecolor{red}{RGB}{255,0,0}

    \begin{axis}[
        width=8cm,
        height=6cm, 
        xlabel={$k$},
        ylabel={$x$},
        xmin=0, xmax=10,
        ymin=0, ymax=6,
        xtick={0, 2,...,10},
        ytick={0,1,...,6},
        tick label style={font=\footnotesize},
        axis x line=bottom,
        axis y line=left,
        axis on top=true,
        legend style={
            at={(0.5, 0.97)}, 
            font=\scriptsize, 
            anchor=north, 
            legend columns=1, 
            draw=none, 
            fill=white, 
            fill opacity=0.5, 
            text opacity=1, 
            rounded corners=2pt,
            row sep=-2pt,
            cells={anchor=west, align=left}
        },
        x label style={at={(axis description cs: 1.01, -0.0)}, anchor=west},
        y label style={at={(axis description cs: -0.08, 0.5)}, anchor=east},
    ]

    \addlegendimage{line legend, line width=1pt, color=blue}
    \addlegendentry{Iteration $0$}
    
    \addlegendimage{line legend, line width=1pt, color=green}
    \addlegendentry{Iteration $3$}
    
    \addlegendimage{line legend, line width=1pt, color=orange}
    \addlegendentry{Iteration $6$}
    
    \addlegendimage{line legend, line width=1pt, color=purple}
    \addlegendentry{Iteration $9$}
    
    \addlegendimage{line legend, line width=1pt, color=red}
    \addlegendentry{STL-OCP solution}
    
    \addplot graphics [xmin=0, xmax=10, ymin=0, ymax=6] {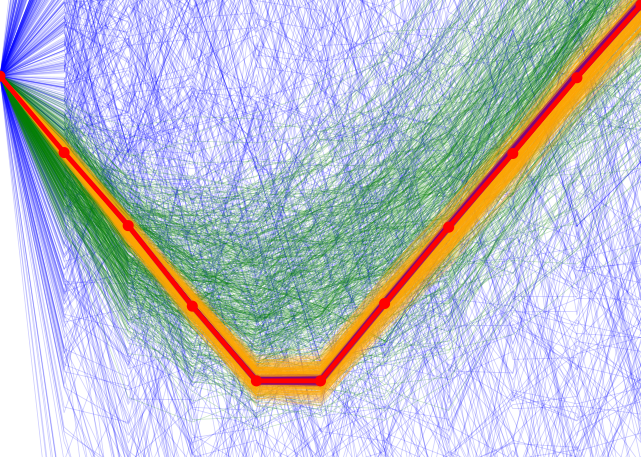};
    
    \end{axis}
\end{tikzpicture}}
    \vspace*{-3mm}
    \caption{STL-OCP solution of our PI solver for problem I. Also, the sampled trajectories of four iterations are presented.}
    \label{fig:SolutionSimpleExample}
\end{figure}

The solution of our \textbf{PI} solver is presented in Fig.~\ref{fig:SolutionSimpleExample}. It satisfies $\varphi_1$ and $x_K$ reaches a maximum. Fig.~\ref{fig:SolutionSimpleExample} visualizes the sampled trajectories for four iterations, illustrating the shrinking variance of the samples over the iterations.
Table~\ref{tab:PerformanceResults} presents the optimal costs $C^*$, the robustness $\rho^{\varphi}$, and the convergence time $t_{\mathrm{c}}^*$ for the solutions of the different solvers. All solvers, except the \textbf{GRAD} solver, provide similar optimal costs and do not violate $\varphi_1$. 


\subsection{Problem II: Simple Motion Planning}
We now revise the motion planning problem from Fig.~\ref{fig:Initial_Example}, where the goal is to eventually reach the blue rectangular area while avoiding a collision with the gray circular obstacle in the planning time horizon ${K=15}$. A state is defined as $x:= [\mathtt{p}_x, \mathtt{p}_y, \mathtt{v}_x, \mathtt{v}_y]$, where $\mathtt{p}_x, \mathtt{v}_x \in \mathbb{R}$ are the position and the velocity in $x$-direction, respectively, and $\mathtt{p}_y, \mathtt{v}_y \in \mathbb{R}$ are the position and the velocity in $y$-direction, respectively. The input is $u:= [\mathtt{a}_x, \mathtt{a}_y]$, where $\mathtt{a}_x, \mathtt{a}_y \in \mathbb{R}$ are the accelerations in $x$-direction and $y$-direction. We use standard discrete-time double integrator dynamics.
Using STL, the objective is formalized as follows:
\begin{align*}
    \varphi_2 &:= G_{[0,K]} (\lnot\mu_{\text{in\_circle}}) \land F_{[0,K]} (\varphi_{\text{in\_box}}), \text{ with}\\
    \varphi_{\text{in\_box}} &:= \mu_{\text{above\_of}} \land \mu_{\text{below\_of}} \land \mu_{\text{right\_of}} \land \mu_{\text{left\_of}},
\end{align*}
and the predicates are:  
\begin{align*}
    \mu_{\text{in\_circle}}&:= r^2 - (\proj_{\mathtt{p}_{x}}(x_k) - m_x)^2\\
                                     &\phantom{:= r^2\ } - (\proj_{\mathtt{p}_{y}}(x_k) - m_y)^2 \geq 0,\\
    \mu_{\text{above\_of}} &:= \proj_{\mathtt{p}_{y}}(x_k) - \underline{\mathtt{p}}_{y} \geq 0,\\
    \mu_{\text{below\_of}} &:= \overline{\mathtt{p}}_{y} - \proj_{\mathtt{p}_{y}}(x_k)  \geq 0,\\
    \mu_{\text{right\_of}} &:= \proj_{\mathtt{p}_{x}}(x_k) - \underline{\mathtt{p}}_{x} \geq 0,\\
    \mu_{\text{left\_of}}  &:= \overline{\mathtt{p}}_{x} - \proj_{\mathtt{p}_{x}}(x_k)  \geq 0,
\end{align*}
where $r, m_x, m_y, \underline{\mathtt{p}}_{x}, \overline{\mathtt{p}}_{x}, \underline{\mathtt{p}}_{y}, \overline{\mathtt{p}}_{y} \in \mathbb{R}$ are parameters. Due to $\mu_{\text{in\_circle}}$, problem II is nonlinear and has two distinct optimal solutions. The terminal costs are $E(x_K):= \proj_{\mathtt{p}_{x}}(x_K) + \proj_{\mathtt{p}_{y}}(x_K)$.

Again, a numerical comparison is provided in Tab~\ref{tab:PerformanceResults}. All solvers find a solution that satisfies $\varphi_2$, and they achieve similar optimal costs. The increased complexity of problem~II, compared to problem~I, leads to a higher convergence time for all solvers besides our \textbf{PI} solver. 
The increased computation time for the \textbf{CMA-ES} solver likely indicates that this covariance-adaption approach has more difficulty deciding on one of the two possible solutions. Using a smooth STL cost term, the \textbf{SGRAD} solver converges faster than the \textbf{GRAD}~solver.


\subsection{Problem III: Complex Motion Planning}
Finally, we present a challenging motion planning problem. The state is defined as $x:= [\mathtt{p}_x, \mathtt{p}_y, \theta, \mathtt{v}, \Psi]^\top$, where $\mathtt{p}_x, \mathtt{p}_y$ are the positions in $x$-direction and $y$-direction, respectively, $\theta$ is the steering angle, $\mathtt{v}$ is the velocity, and $\Psi$ is the orientation. The input is $u:= \left[\mathtt{v}^{\theta}, \mathtt{a}\right]^\top$, where $\mathtt{v}^{\theta}$ is the steering velocity and $\mathtt{a}$ is the longitudinal acceleration. We consider a kinematic single-track model with an explicit Euler integration as: $x_{k+1} = x_k + \check f({x}_k,u_k) \Delta t,$ with 
\begin{align*}
    \check f({x}_k,u_k)   &= \left[\mathtt{v}_k \cos(\Psi_k), \mathtt{v}_k \sin(\Psi_k), \mathtt{v}^{\theta}_k, \mathtt{a}_k, \frac{\mathtt{v}_k}{\ell} \tan(\theta_k)\right]^\top,
\end{align*}
where $\ell \in \mathbb{R}$ is the wheelbase of the vehicle. The planning time horizon is $K=50$.

We intend the system to a) remain in the area $\alpha$, to b) not collide with any of the five obstacles $\mathcal{O}$, to c) stay at least for two consecutive time steps in each of the task areas $\tau \in \mathcal{T} := \{\tau_1, \tau_2, \tau_3\}$ (while they can be visited in arbitrary order), and to d) not be simultaneously at task area $\tau_1$ and $\tau_2$. In STL, this is formalized as follows:
\begin{align*}
    \varphi_3 := \varphi_{\text{in\_area}} \land  &\varphi_{\text{no\_collision}} \land \varphi_{\text{do\_tasks}} \land \varphi_{\text{avoid\_overlap}}, \text{ with}\\
    \varphi_{\text{in\_area}} &:= G_{[0, K]}\left(\varphi_{\text{in\_box}}^\alpha\right),\\
    \varphi_{\text{no\_collision}} &:= G_{[0, K]}\left(\textstyle\bigwedge_{o \in \mathcal{O}} (\lnot \varphi_{\text{in\_box}}^o)\right),\\
    \varphi_{\text{do\_tasks}} &:= \textstyle\bigwedge_{\tau \in \mathcal{T}}\left(F_{[0, K]}\left(G_{[0, 1]}(\mu_{\text{in\_circle}}^\tau)\right)\right),\\
    \varphi_{\text{avoid\_overlap}} &:= G_{[0, K]}(\lnot(\mu_{\text{in\_circle}}^{\tau_1} \land \mu_{\text{in\_circle}}^{\tau_2})).
\end{align*}
The superscripts $\alpha$, $o$, and $\tau$ indicate that the required parameters are provided to the respective subformulas and predicates. The terminal costs are $E(x_K):= \proj_{\mathtt{p}_{x}}(x_K) + \proj_{\mathtt{p}_{y}}(x_K)$.

\begin{figure}[!tb]
    \centering
    \vspace*{2mm}
    \resizebox{0.95\columnwidth}{!}{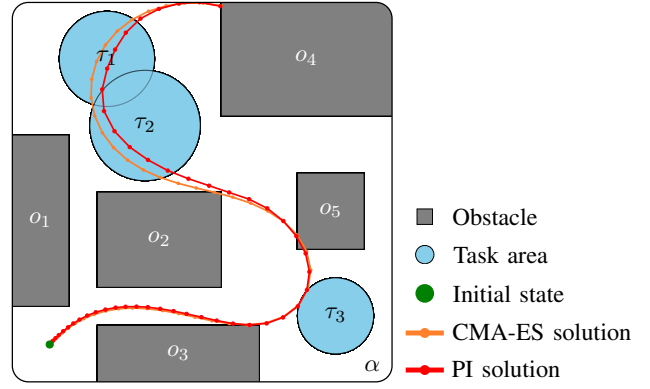}
    \vspace*{-3mm}
    \caption{Solutions for problem III. Note that we evaluate the robustness function only at discrete time steps.}
    \label{fig:SolutionComplexExample}
    \vspace*{-5mm}
\end{figure}

The \textbf{PI} solver converged in \SI{25.07}{\second}, and it took \SI{79.82}{\second} for the \textbf{CMA-ES} solver. Their solutions satisfy $\varphi_3$ and are presented in Fig.~\ref{fig:SolutionComplexExample}. However, the remaining solvers did not converge within a limit of \SI{3600}{\second}. This indicates the capability of the sampling-based methods, especially of our PI approach, to solve complex STL-OCPs in adequate time without the need to reformulate the robustness term. The performance of the presented \textbf{PI} solver can be even more improved by utilizing a GPU. However, a comprehensive numerical benchmark with many problems and further state-of-the-art solvers is essential for future work.

\section{Conclusions}\label{sec:Conclusions}
This paper introduces a novel sampling-based solution method for deterministic STL-OCPs within the PI framework. A key feature of this sampling-based approach is the gradual reduction of the problem variance, which enables the handling of  non-differentiable problem formulations typically associated with STL. Additionally, the method maintains MPPI’s inherent advantage of a straightforward implementation and parallel execution, making it suitable for real-time applications. Through numerical experiments, the performance of the method is demonstrated and compared to other state-of-the-art methods for solving STL-OCPs. The findings suggest that our approach holds promise for advancing the design of autonomous systems by efficiently addressing complex planning problems.

Future work will focus on extensively comparing the performance of our method with other state-of-the-art solvers and enhancing it to perform model predictive control with STL specifications. Additionally, we intend to extend the method to handle prioritized STL specifications.

    \section*{Acknowledgment}
The authors gratefully acknowledge financial support from the German Research Foundation (DFG) under grant AL 1185/19-1 and thank Moritz Diehl, Samuel Henkel, and Florian Messerer for the valuable discussions that contributed strongly to this paper.

	\bibliographystyle{IEEEtran}

\end{document}